\documentstyle[12pt]{article}
\begin{document}
\
\vskip 4truecm

\centerline{{\bf  A Note on Spontaneous Baryogenesis  }}

\centerline{{\bf   Michael Joyce}}

\bigskip

\centerline{\it {Joseph Henry Laboratories,}}

\centerline{\it {Princeton University, }}

\centerline{\it {Princeton, NJ O8544.}}

\bigskip

\bigskip

\bigskip

\centerline{{\bf {\it Based on a talk given  at the conference on}}}
 \centerline{{\bf{\it ELECTROWEAK PHYSICS AND THE EARLY UNIVERSE}}}
 \centerline{{\bf{\it held at Sintra, Portugal from March 22 to 27, 1994.}}}

\bigskip
\bigskip
\bigskip

May 1994.

\centerline{{\bf Abstract }}
\bigskip
The original calculation of `spontaneous' baryogenesis overlooked the
role played by transport of particles onto the bubbles wall. For typical
`adiabatic' wall thicknesses and velocities one can model the problem
in a fluid approximation and the mechanism is best understood as another
limit of `charge transport' baryogenesis, and can produce a similar
non-local effect in front of the wall.

\pagebreak

In this brief talk I will make a few comments on `spontaneous
baryogenesis', a mechanism for the production of baryons
at the electroweak
phase transition in the `adiabatic' limit of thick slowly moving bubble walls
proposed by Cohen, Kaplan and Nelson (CKN) in
\cite{ckn}. My comments summarize results of a collaboration with
Tomislav Prokopec and Neil Turok. The central point
is that the constraints imposed on particle numbers
in the thermodynamic equilibrium calculations which this mechanism
involves are inappropriate for the relevant wall thicknesses.
A correct calculation of the baryon production
involves taking into account the
effect on the surrounding plasma of what is effectively a
background axial gauge field on the wall. This calculation
should really be thought of as another limit of  `non-local'
baryogenesis, advocated by CKN in the case of very thin bubble
walls and discussed by my collaborators in other talks at this
conference \cite{nttp}, rather than as a distinct mechanism.

Consider the familiar two Higgs doublet model in which the CP violating
phase $\theta$
changes on the bubble wall during the phase transition. If,
for simplicity,
we couple only one of the two doublets to the fermions through
Yukawa terms, the mass terms which result when the Higgs doublet
acquires a vev contain the (gauge invariant) phase $\theta$ and
may be written (in the appropriate gauge)
\begin{equation}
Yve^{i\theta}\bar{q}_Lq_R + Y've^{-i\theta}\bar{q}'_Lq'_R + h.c.
\end{equation}
where $q$ stands for the charge $\frac{2}{3}$ quarks, $q'$ for
the charge $-\frac{1}{3}$ quarks and the charged leptons and
$Y$ and $v$ for the Yukawa coupling and vev.

The observation made by CKN is that if one does a hypercharge rotation
(chosen because it is anomaly free)
on the fermions to remove the phase $\theta$ and make the mass
terms real, the kinetic terms generate
\begin{equation}
2\partial_{\mu} \theta \Sigma_i y_i \bar{\psi}_i \gamma^{\mu} \psi_i
\label{newterm}
\end{equation}
where the sum is over all the fermions which have hypercharge $y_i$.
If one now considers the case in which the spatial gradients
$\partial_i \theta$ are negligible, (\ref{newterm}) looks like
a chemical potential term for fermionic hypercharge. Because it is not
strictly a chemical potential - it does not arise from a constraint-
CKN refer to it as a `charge potential'.  However if one does
thermodynamic calculations with this Hamiltonian the effect of the
term is just like that of a chemical potential e.g. the number
density $n_i$ of a massless fermion $i$ in the presence of a
chemical potential $\mu_i$ is
\begin{equation}
n_i = \frac{T^2}{6}(\mu_i + 2y_i \dot{\theta}).
\end{equation}

So how does this bias electroweak anomalous processes? As the bubble wall
passes through a region this charge potential is turned on and the
system will approach the local thermal equilibrium attained in this
background. CKN calculate this local equilibrium in \cite{ckn} by
imposing constraints on the quantum numbers in the system which are
conserved by the perturbative processes which are slow relative to the
timescale for the wall to pass by. They do this by introducing
chemical potentials $\mu_A$ to constrain each of these charges
to zero (their global value). Amongst these charges is the baryon
number $B$ (because the sphaleron processes are too slow to reach
equilibrium as the wall passes) and  a corresponding chemical
potential $\mu_B$. For example if we take the conserved charges,
for simplicity, to be $Y$, the total hypercharge, $B$ and $B-L$,
where $L$ is the lepton number, we find
\begin{eqnarray}
Y&\propto 20 \dot{\theta}+(10+n)\mu_Y +8\mu_{B-L}+2 \mu_B=0 \cr
       B-L&\propto 16\dot{\theta} +8\mu_Y+13 \mu_{B-L}+ 4 \mu_B=0 \cr
       B&\propto 2\dot{\theta} +\mu_Y+2 \mu_{B-L}+ 2 \mu_B=0.
\end {eqnarray}
using the relation
$\mu_i=\Sigma_A q^A_i \mu_A$ and $n_i \propto k_i \mu_i$
for the particles which
we assume massless. $k_i$ is a counting factor which is
one for fermions and two for bosons, $q^A_i$ is the $Q_A$ charge
of species $i$, and $n$ is the number of Higgs doublets in the
theory.
Solving, we find $\mu_{B} \propto \dot{\theta} n$.

To find the resulting $B$ violation we use the fact that the
rate of change of a `conserved' charge $X$ due to a
`slow' process is
\begin{equation}
\dot{X}= -  \Gamma\frac{\delta F}{T}\delta X
\label{xdot}
\end{equation}
where $\delta F$ and $\delta X$ are the changes in the free
energy  and in $X$ per process respectively and
$\Gamma$ is its equilibrium rate.
Applying this to the electroweak sphaleron processes we have
\begin{equation}
\dot{B}= - \Gamma_{sph}N_F^2\frac{\mu_B}{T}
\label{bdot}
\end{equation}
where $\Gamma_{sph}$ is the equilibrium electroweak sphaleron rate
ans $N_F$ the number of families.
We conclude that $B$ violation on the wall is directly driven by
$\dot{\theta}$ and the final generated asymmetry is simply calculable
from (\ref{bdot}).

There is a simple explanation of the proportionality of the answer to
the number of Higgs doublets. In the local thermal equilibrium
in which each particle species $S_i$ has chemical potential $\mu_i$
the rate at which any particle density $n_i$ changes due to some
out of equilibrium process $\nu_jS_j \rightarrow 0$ (where, for
example, $A\rightarrow 2B+C$  has $\nu_A=1, \nu_B=-2, \nu_C=-1$)
is, by (\ref{xdot}) above,
\begin{equation}
\dot{n}_i= - \frac{\Gamma}{T}(\Sigma_j\nu_j\mu_j)\nu_i.
\label{ndot}
\end{equation}
It follows from this that the particle densities are only affected
directly when $\dot{\theta}$ is turned on by processes which have
Higgs particles in the ingoing or outgoing state. Any other process obeys
$\Sigma_i\mu_i\nu_i=0$ simply by conservation of fermionic hypercharge,
and therefore by (\ref{ndot}) is not out of equilibrium when $\dot{\theta}$
is turned on. In CKN's calculation it is the top/Higgs coupling which
is taken to be `fast'. Forcing this process to be in equilibrium
alters the abundances in such a way as to produce a non-zero rate for the out
of equilibrium sphaleron processes.

It is now instructive to compare this with the case of a  potential $\phi_Y(x)$
for ${\it total}$ hypercharge which is turned on in some region of space.
When such a potential is turned on no process is out of equilibrium
and hence no density directly affected, simply because total hypercharge
is conserved in all processes.  However this does not mean that the
densities of particles do not change if the potential is turned on
in some small region. In fact the potential will be screened by
drawing in charge from the plasma. The local equilibrium, which is
a solution of the Boltzmann equation, is given by the particle phase
space distributions $f_i$
\begin{equation}
f_i(p,x) = \frac {1}{e^{\beta E_i(x)} \pm 1}
\label{eqm}
\end{equation}
where $E_i=\sqrt{p^2 + m_i^2}+y_i\phi_Y(x)$ and, therefore,
$n_i \propto y_i \phi_Y(x)$.

In the case of the charge potential $\dot{\theta}$ there is also
precisely such an unconstrained equilibrium solution, except that
$E_i$ is now the energy level of species $S_i$ in the presence of
$\dot{\theta}$. If this equilibrium is attained neither the Higgs
processes nor the sphaleron processes are out of equilibrium. In the
same way as in the presence of a total hypercharge potential
there can be transport processes which bring the system towards
the static equilibrium (\ref{eqm}).

So the central question is whether there can be significant transport of
charge into the small region (the bubble wall during the first order
phase transition) over the relevant timescale.
A simple estimate is as follows. The distance a particle with
diffusion constant $D$ typically
diffuses in a given direction
in time $t$ is $\sqrt{2Dt}$. In order that the transport
processes be negligible
the distance a particle diffuses in the time the wall of length $L$
moving with velocity $v_w$ takes to pass
should be much less than the wall thickness i.e.
\begin{equation}
L >> \frac{2D}{v_w}.
\label {cond}
\end{equation}

For quarks $D \sim \frac{10}{T}$ and in order to allow  the
Higgs/top processes, for which $\Gamma \sim  \frac{T}{50}$,
time to equilibrate we need a small $v_w$ for a typical
`thick' wall with $L \sim \frac {20 - 40}{T}$. For typical
`adiabatic' wall thicknesses and velocities therefore
the constrained equilibrium calculated
by CKN is not attained.

A more quantitative analysis requires a full treatment which we
will present elsewhere \cite{jpt}, but bears out this rough
qualitative argument. The problem becomes that of
the determination of the deviations caused by the motion of the
wall from the equilibrium (\ref{eqm}) when $\frac{v_w L}{D} < 1$.
Given that the wall is thicker than the mean free path of the
quarks it is valid to treat the problem in
a fluid approximation. The effect of the charge potential
$\dot{\theta}$  - or more generally of $\partial_{\mu}\theta$ -
is to both alter the rates of processes coupling the fluids
locally ${\it and}$ to induce a force pulling or pushing the
fluids around on the wall.

To conclude we return to the lagrangian after the
hypercharge rotation which can be written
\begin{equation}
\bar{\psi}(i\partial+\frac{y_L+y_R}{2}\partial \theta
                    -\frac{y_L-y_R}{2}\gamma^{5} \partial\theta)\psi
                    +m\bar{\psi}\psi.
\label{vaform}
\end{equation}

The vector part of the induced term can be rotated away by making use of the
surviving electromagnetic vector $U(1)$
\footnote{The corresponding ambiguity about
the original rotation was ignored in our discussion above.}. One can see
explicitly that the effect of the changing phase is to turn
on an axial $U(1)$ gauge field, which despite the fact that it is pure
gauge cannot be removed because the axial symmetry is explicitly broken
by the mass term.

The physical effects of this field can be understood
in various different limits. In fact the calculation of the solutions for
a Dirac particle in this background is what is done to find the
reflection coefficients for the different chirality quarks or
leptons in the `charge transport' mechanism. In the present
case where the wall is thick and the
particles cannot be treated as free we need to understand the
dynamics of the particles while they are on the wall. Working
in the WKB limit (which should be good for most of the particles
on the wall if we take $\partial_z \theta \sim L^{-1} << T$ )
one can calculate an effective chiral `classical'
force, which one then uses in the fluid equations. This can
lead to a chiral disturbance {\it in  front  of}  the bubble wall, just
as in the `charge transport' mechanism. In contrast
to that case the chiral asymmetry in front of the wall is
not driven by a reflected flux
of particles with $p_z \sim L^{-1}$ but by the effect on the fluid
of a build-up of particle densities on the wall
due to a classical force which acts on particles of {\it all momenta}
as they cross the wall. Despite the fact that the reflection is very
suppressed for thick walls
the same $\partial_{\mu}\theta$ field can in this case
produce a chiral asymmetry
and `non-local' baryogenesis in front of the wall through its effect on
a different part of the phase space.
We will explore this limit of
`classical non-local' baryogenesis in a forthcoming publication.

\bigskip
\centerline {\bf Acknowledgements}

In addition to my collaborators T. Prokopec and N. Turok, I am grateful
to S. Nasser and M. Shaposhnikov for useful discussions. This
work was done with the support of a Charlotte Elizabeth Procter Fellowship
at Princeton University.

\end{document}